# Surface Sputtering from Cold Dark Matter Interactions: Proposed Search for its Diurnal Modulation.

J. I. Collar[§] and F. T. Avignone, III

Department of Physics and Astronomy
University of South Carolina
Columbia, SC 29208

Nuclear recoil cascades induced by Cold Dark Matter (CDM) elastic scattering can produce the ejection of target atoms from solid surfaces. We calculate the yield and energy distribution of these sputtered atoms in a variety of materials. These parameters would suffer a large diurnal modulation induced by the rotation of the Earth and its motion through the galactic halo. Schemes for the detection of this unique CDM signature are proposed.

[§] Corresponding author. e-mail: ji.collar@scarolina.edu

1.  **INTRODUCTION**

Approximately 90% of the matter in the Universe eludes detection by means of conventional astronomy; the problem of determining the nature of this dark matter has existed for the last sixty years. During this period, indirect evidence for its existence has mounted [1]. Numerous candidates have been proposed, ranging from familiar baryonic dark matter (BDM), to massive neutrinos and more exotic particles and possibilities. Cold Dark Matter (CDM), consisting of heavy, non-relativistic particles, extends over a large range of possible masses and coupling constants to conventional matter. Weakly Interacting Massive Particles (WIMPs) belong to this group and comprise some of the candidate particles more likely to constitute the galactic dark halo. The ample observational evidence for this halo is heralded by the recent studies of the dynamics of the Large Magellanic Cloud [2]. The theoretical justification for WIMPs stems from extensions of the Standard Model of the electro-weak interaction. In some cases, WIMPs have the added attraction of having been initially postulated to solve theoretical problems unrelated to that of Dark Matter, as in the case of the Lightest Supersymmetric Partner, LSP, (a neutralino) [3].

The WIMP hypothesis has prompted the development of new experimental techniques aimed at the direct detection of recoil energy from WIMP coherent elastic scattering off nuclei [4]. The small rates of interaction make separating the CDM signal from backgrounds the main experimental challenge. Operational detectors may already be detecting these WIMP events, but the inability to isolate this small signal would prevent one from making any kind of identification. Phonon-sensitive detectors [5] promise nearly perfect discrimination between nuclear recoil events and ionizing radiation. Even in this ideal case, processes such as neutron elastic scattering can mimic the WIMP signal. A reduction in the background would, *per se*, only be able to exclude some candidates but not to make a discovery. To achieve this, one must observe signatures that are unique to the WIMP-nucleus interaction and that convey information about the CDM particle, such as its mass and elastic cross section. At least two such signatures, modulations in the rates and kinetic energies of the nuclear recoil signals, have been proposed. Both exploit the present understanding of the halo dynamics and of the Earth's movement through it. The first approach, referred to as "yearly modulation" [6,7], takes advantage of the variation in the vectorial sum of the Earth's orbital velocity and that of the Sun through the halo. The second proposal [8] relies on local daily changes in CDM flux and speed distribution, induced by the elastic scattering of CDM particles in the Earth's interior. The annual character of the first method imposes strict demands on the stability of the detector and length of the search; it is also arduous to separate this modulation from variations due to long-lived backgrounds. The yearly modulation is small ($\pm 4.5\%$ annual change in the mean energy deposition and $\pm 2.5\%$ in the overall detection rates, requiring massive detectors for its observation) but, due to its purely kinematic origin, should be present for any detectable CDM particles comprising the halo. The daily periodicity of the second method increases its statistical effectiveness, but it has the clear disadvantage of being dependent on the CDM particle's elastic cross section and mass, making the modulation extremely small in the case of most neutralino matter. Further investigation and refinement are obviously needed in this area.

Here we describe a measurable process that undergoes a large diurnal modulation and applies to all detectable CDM candidates. Their elastic scattering results in a primary recoiling nucleus that loses its energy through ionization and further nuclear collisions that produce a cascade of secondary recoils. If the interaction takes place close to the surface of a solid, the recoiling cascade can be partially expelled and subsequently detected (delta electrons from ionization losses have much shorter ranges, making their ejection comparatively infrequent). The sputtering yield and energy distribution of these ejected atoms undergo a strong daily variation. The origin of this modulation can be explained in three steps: first, the movement of the Earth through the halo (where the CDM velocity distribution is isotropic), boosts the CDM velocities in our reference frame, creating a preferred direction (i.e., a galactic "wind").  Second, this directionality of the CDM flux is preserved to some extent by the kinematics of the WIMP-nucleus interaction, being transferred to the recoils. Third, and most importantly, the directionality is partially maintained by the straggling of the cascade atoms. The Earth's rotation changes the preferred direction relative to the detector, producing the daily variations. We follow this chronology of events in this paper.

## 2. THE GALACTIC "WIND"

CDM particles form an spherical, dissipationless dark halo (generally assumed to be non-rotating), with an isotropic speed distribution in the galactic rest frame given by [7]:

$$p(v)dv \propto v^2 \cdot \exp\left(\frac{-3v^2}{2v_{dis}^2}\right) dv \quad , \qquad (1)$$

where $v_{dis} = 270 \pm 25$ km/s is the dispersion in the CDM speeds. This distribution is truncated at the local galactic escape velocity $v_{esc} = 500 - 650$ km/s. A simple Galilean transformation converts the distribution to the Earth's reference frame, where it takes the form [7]:

$$f(v)dv \propto X^2 \cdot \exp\left[-\left(X^2 + \eta^2\right)\right] \cdot \frac{\sinh(2X\eta)}{2X\eta} dX \quad . \qquad (2)$$

Here X is a dimensionless CDM speed with respect to Earth and $\eta$ is the dimensionless speed of the Earth through the halo:

$$X^2 = \frac{3v^2}{2v_{dis}^2} \quad ; \quad \eta^2 = \frac{3v_{Earth}^2}{2v_{dis}^2} \quad . \qquad (3)$$

The value of $v_{Earth}$ (annual average ~ 260 km/s) is derived from a detailed analysis of the Earth's movement through the galaxy, discussed below.
For an Earth-bound observer, the similar magnitude of $v_{Earth}$ and $v_{dis}$ results in a strong directionality of the CDM velocity vectors in the direction opposite to $\mathbf{v}_{Earth}$, and in an azymuthal symmetry around this axis. Let us define a polar angle $\theta$ between any given direction in space and this preferred one, $\mathbf{w} \equiv -\mathbf{v}_{Earth}$. Figure (1) displays the distribution of CDM velocity vectors in the Earth's frame, as a function of $\theta$ and their magnitude. It is observed that ~ 95 % of the CDM particles travel in directions $\theta < 90°$. These have a mean speed of $< v > \sim 350$ km/s, while $< v > \sim 200$ km/s for those with $\theta > 90°$. We shall refer to $\mathbf{w}$ as the direction of this galactic "wind".

Let us consider a terrestrially stationary target material, with a flat surface defined by its normal vector, **r**, expressed in a geocentric coordinate system. A suitable choice of astronomical coordinates, the Equatorial System [9], uses the axis of rotation of the Earth and the position of the Sun in the sky at the Vernal Equinox (VE) as reference directions. It is relatively simple to express **r** and the Earth's orbital velocity vector ($v_{Earth}^{orb}$ ~ 30 km/s around the Sun) in the equatorial coordinates $\alpha$ ("right ascension") and $\delta$ ("declination"), and to observe the time evolution of their coordinates [10]. At the moment of the VE, $\mathbf{v}_{Earth}^{orb}$ points in the direction $\alpha = 270°$, $\delta = -23.44°$, revolving counterclockwise around the ecliptic and completing a revolution in a tropical year (365.2422 solar days). The coordinates of **r** depend on the geographical location of the target material and its orientation, and change daily with the Earth's rotation.

The largest component of $\mathbf{v}_{Earth}$ comes from the translation of the Sun around the galactic center in the galactic plane or disk, at $v_{Sun}^{disk} = 250 \pm 25$ km/s [9]. There is a small peculiar component of the Sun's velocity off the plane of the galaxy ($v_{Sun}^{pec}$ ~ 16.5 km/s) that is also included in our calculations. Both are best expressed in the Galactic Coordinate System [9]. This system takes the plane of the Galaxy as its fundamental plane and the line joining the Sun to the center of the Galaxy as the reference direction. The transformation between Equatorial and Galactic coordinates found in the Ephemeris and Nautical Almanacs is rather cumbersome, but convenient computer algorithms exist [11]. Once $\mathbf{v}_{Earth}^{orb}$ and **r** have been translated to galactic coordinates, the three components of the Earth's net velocity through the halo ($\mathbf{v}_{Earth}^{orb}$, $\mathbf{v}_{Sun}^{disk}$ and $\mathbf{v}_{Sun}^{pec}$) can be vectorially added. The resulting vector $\mathbf{w} \equiv -\mathbf{v}_{Earth}$ and **r** are now defined in the same coordinate system and the angle $\theta$ between them can be computed for any given time, detector location and orientation. Following this procedure, one can obtain the yearly modulation in $v_{Earth}$ of Ref. [6] as a cross-check.

The calculation of $\theta$ becomes especially simple if **r** is aligned with the radius vector from the center of the Earth to the location of the target material, i.e., when the target's planar surface is parallel to the ground. Fig. (2) shows the dependence of $\theta$ on the geographical location of the material and time of the day. The daily maxima in $\theta$ for a given location change only slightly through the year (~10 %), a result of the orbiting motion of the Earth.

For this choice of target orientation, the geographical latitude defines the amplitude of the variation; maximal fluctuations occur at 35°-50° North and South latitudes.

## 3. KINEMATICS OF THE WIMP-NUCLEUS INTERACTION

Consider a thin planar target oriented perpendicular to $\mathbf{w}$, such that $\theta = 180°$ for the normal vector to the "upwind" surface. Many more CDM particles will enter the material through this side than through the "down-wind" surface. The converse applies to particles exiting the material. Also, those leaving (entering) the "down-wind" ("upwind") surface are generally more energetic (see Fig. 1). If at a later time the rotation of the Earth makes the sheet parallel to $\mathbf{w}$, the number and energy distributions of entering and exiting particles are equal and the same for both surfaces.

This variation in the number and energies of particles leaving a target's surface, immediately suggests a possible similar change in the yield and energy distribution of sputtered atoms arising from their interactions. In order to establish this possible link we must define the relationship between the initial direction of a CDM particle and that of the recoiling nucleus that it may produce. For a CDM particle of mass $m_\delta$ and speed $v$, and a target nucleus of mass M initially at rest, the initially available kinetic energy is $T_0$. We denote the kinetic energy of the CDM projectile after elastic scattering as T' and that of the recoiling nucleus as T. The scattering angle of $m_\delta$ is $\psi$, while M recoils at an angle $\xi$ from the initial CDM trajectory. Conservation of momentum dictates $\xi \leq \pi/2$, which itself guarantees that the anisotropy of the CDM flux is at least partially passed on to the recoils; since most CDM particles travel in $\theta < 90°$ directions, most recoils will have initial velocity vectors pointing in the same hemisphere. In particular, if $m_\delta > M$, the upper limit of $\psi$ is $\psi_{max} = \arcsin(M/m_\delta)$, with the largest energy transfer occurring at this limiting angle. Conservation of energy and the relation [12],

$$\frac{T'}{T_0} = \frac{m_\delta^2}{(m_\delta + M)^2} \left[ \cos\psi \pm \sqrt{\left(\frac{M}{m_\delta}\right)^2 - \sin^2\psi} \right]^2 \quad , \tag{4}$$

define the maximum recoil energy in a WIMP-nucleus interaction, $T_{max} = M v^2$, which takes place in the limit $m_\delta >> M$, when $\psi$ is necessarily small and $\xi$ peaks close to $\pi/2$. We expect in this limit an enhancement in the number of nuclear recoils in the general direction of $\theta \sim 90°$. For smaller values of $m_\delta$, the distribution of recoils is more diffuse, but it is always largest in the forward direction ($\theta < 90°$), due to the condition $\xi \leq \pi/2$.

## 4. TRANSPORT OF RECOILING IONS

The recoiling nucleus loses energy discretely through nuclear collisions and continuously in electronic interactions. Both mechanisms have been extensively studied for a wide range of ion energies and materials [13,14]. Even though the spatial anisotropy of the CDM flux is reflected in the spatial distribution of primary recoiling nuclei, one might expect that their subsequent interactions would tend to diminish any directional effects. The initial direction of an ion penetrating a solid defines a plane perpendicular to it. The energy-reflection coefficient [15] is the fraction of its kinetic energy that is reflected back behind this plane, be it in the form of delta-electrons, a backscattered primary ion, or secondary ions from the cascade of nuclear recoils characteristic of nuclear stopping. In the case of low energy self-irradiation (target atoms recoiling through the target material), this coefficient is less than $\sim 5\%$ for all elements [15], indicating that most energy is dissipated in the initial direction of the primary recoil. This completes the description of the process by which the anisotropy in the CDM flux is transmitted to the cascade of nuclear recoils. Accordingly, we expect a dependence of the characteristics of the sputtered species on the orientation of the target material with respect to the galactic "wind".

In order to calculate this dependence we have performed Monte Carlo calculations for a variety of elements, spanning the periodic table. Using heavy neutrinos as the CDM particle, we simulate their interactions in thin sheets of target materials, assuming a specific orientation of the sheet with respect to $\mathbf{w} \equiv -\mathbf{v}_{Earth}$. The direction of each recoil is obtained, and their transport followed using the 1992 version of the widely used computer code TRIM (TRansport of Ions in Materials) [13,16]. This code has a documented accuracy of a few percent in the low energy region ( $T \sim$ few keV ). TRIM employs various theoretical models for the interatomic potentials that define nuclear scattering. The electronic losses are computed using the velocity dependent treatment of Lindhard-Scharff [17] or Bethe-Bloch theory, depending on the ion's energy. The code follows primary and secondary ions in their straggling through the material, and yields precise information about the ejected atoms (namely their individual energies, exit points and directions). The predictions of TRIM in sputtering studies are in excellent agreement with experiment [18].

## 5. MONTE CARLO SIMULATIONS AND NUMERICAL RESULTS

We begin a simulation by defining a spatial direction for the vector **r**, the normal to a thin sheet of target material. We have selected θ = 0°, 90°, 180° as representative orientations. In the calculations we consider sputtering from one single side of this sheet, namely the one from which **r** stems. TRIM provides the mean projected range, $R_P(T_{max})$, corresponding to a recoiling nucleus with $T_{max} = M v_{max}^2$. The speed $v_{max}$ is the maximum for a WIMP relative to Earth ($v_{max} = v_{Earth} + v_{esc}$). The thickness of the sheet is chosen to be $\lambda = R_P(T_{max})$ for every material, typically 700 - 1700 Å. For a larger $\lambda$, the contribution to sputtering from nuclei beyond this range is entirely negligible. The surface area of the sheet is kept constant (= 1 m$^2$).

Next, a large number, J, of interaction sites are randomly distributed in the material. The initial trajectory and speed v of the WIMP projectile is selected from the distribution depicted in Fig. (1). The recoil energy T transferred to the nucleus is determined from the differential cross section of the particular CDM candidate under consideration. We digress briefly to explain this point.

The WIMP-nucleus differential rate of scattering is given by [19]:

$$\frac{dR}{dT} = K \frac{\rho_{halo}}{m_\delta} \int_0^{v_{max}} f(v) v \frac{d\sigma}{dT} dv, \qquad (5)$$

where K is the number of target nuclei, $\rho_{halo}$ is the local halo density (taken here $\rho_{halo} = 0.4$ GeV /c$^2$/ cm$^3$ [1]), and f(v) is the normalized WIMP speed distribution given by Eq. (2). We have chosen heavy neutrinos as the WIMPs in our calculations, with the elastic scattering differential cross section [4]:

$$\frac{d\sigma}{dT} = \frac{G_F^2}{8\pi} \cdot \left(\frac{G_f}{G_w}\right)^2 \cdot \left(N - \left(1 - 4 \cdot \sin^2\theta_w\right) \cdot Z\right)^2 \cdot \frac{m_R^2}{T_{max}(m_\delta)} \cdot F(q^2). \quad (6)$$

In equation (6), $m_R$ is the reduced mass, $T_{max}(m_\delta)$ is the maximum recoil energy for a given $m_\delta$, N and Z are the number of neutrons and protons in the target nucleus, and $\theta_w$ is the weak mixing angle ($\sin^2\theta_w = 0.2259$). $G_F^2$ is the Fermi weak coupling constant ($G_F^2 \cong [290 \text{ GeV}]^{-4} = 5.24 \cdot 10^{-38} \text{ cm}^2$) and the parameter $(G_f/G_w)^2$ allows for coupling constants different from $G_F^2$; $G_f < G_w$ for sub-$Z^\circ$ couplings and $G_f = G_w$ for heavy Dirac neutrinos. The term $F(q^2)$ is a form factor accounting for the loss of coherence for very massive projectiles, where forward scattering and lower values of T are favored. We have tested different expressions for $F(q^2)$ and found that the exponential approximation of Ref. [19],

$$F(q^2) = \exp\left(\frac{-8 \cdot \pi^2 \cdot M \cdot T \cdot \varepsilon^2}{3 \cdot h^2}\right), \quad (7)$$

is adequate even for the heaviest nuclei [20] ($\varepsilon$ is the nuclear radius and h is Planck's constant). The selection of heavy neutrinos as the CDM particles is justified by the simplicity in the computation of their rates of interaction and energy transfer to the nucleus. While neutralinos ($\chi^0$) are of great theoretical interest, their cross sections are more parameter-dependent and not yet definitive; revisions have been made recently [21]. Moreover, the coherent spin-independent mode of interaction from Higgs boson exchange [22] ($\sigma_{COH}^\chi \sim m_R^2(N+Z)^2$) prevails over spin-dependent channels for most neutralinos with a zino-higgsino mixture [23,24]; in the low-energy approximation their differential cross-section depends on T only through the same form factor of Eq. (7) [20,24]. In this simple case, $F(q^2)$ alone defines the spatial and energy distributions of recoiling nuclei. The collision kinematics for this important neutralino sector and those of heavy neutrinos are then the same to a good approximation. As a result, sputtering yields for these neutralinos can be derived from those presented below for heavy Dirac neutrinos ($\nu_D$). They scale as the ratio of their respective total rates of interaction in the target material. Our results for energy distributions of the sputtered species apply to this neutralino sector, without modification. For instance, assuming the minimal supersymmetric model and a 40 GeV / $c^2$ Higgs boson [25], one obtains for the ratio of total interaction rates in germanium:

$$\frac{R_{\chi^0}}{R_{\nu_D}} < \text{Exp}\left(773.81 - 999.92\eta + 503.04\eta^2 - 124.21\eta^3 + 15.075\eta^4 - 0.72081\eta^5\right)$$

, (8)

where $\eta = \ln(m_\delta [\text{GeV} / c^2])$. The expression is valid for the range of neutralino masses 20 - 200 GeV / $c^2$, which is theoretically favored [25].

Returning to the simulation, T is randomized according to Eq. (6) for each interaction, and the direction of the recoil obtained as follows [12]:

$$\xi = \arccos\left(\sqrt{\frac{T M}{2 v^2 m_R^2}}\right). \qquad (9)$$

This information is transferred to TRIM, which follows the primary and secondary (cascade) nuclei until all energy is dissipated. In the event that they reach the surface, TRIM registers their location, direction and energy. Typically, a sputtering event consists of one or two energetic atoms (Kinetic Energy ~ few keV) accompanied by a low-velocity cluster (~ 10 atoms, KE ~ few eV). Since their ejection points are separated by only a few angstroms, we define the energy of the event, $E_s$, as the sum of all kinetic energies.

To obtain the sputtering yield, Y, we first compute the total rate of interaction in the material under study, expressed in interactions / kg / year, by integration of Eq. (5). This rate is then multiplied by the mass of the sheet and the sputtering efficiency to obtain Y in units of sputtering events / $m^2$/ year. The sputtering efficiency is simply defined as the number of sputtering events in the run, divided by the size (J) of the simulation. Increasing the thickness ($\lambda > R_P(T_{max})$) does not alter the results, while it diminishes the effectiveness of the simulation by placing too many interactions deep into the material, from where sputtering is energetically forbidden. A smaller $\lambda$ can bias the energy distribution of the ejected atoms towards larger values of $<E_s>$.

It is not straightforward to predict the dependence of Y on the atomic number Z. While the coherence factor in Eq. (6) ($\sim N^2$) favors heavier elements, the projected range for self-irradiation decreases rapidly with atomic mass (and shows some structure in Z). This diminishes the thickness of the active layer that contributes to the sputtering process. Results are shown in Fig. (3a), where $Y(Z,\theta)$ is presented for a 100 GeV/$c^2$ Dirac neutrino. Y is roughly linear with Z, while the amplitude of its modulation between $\theta=0°$ and $\theta=180°$ can be as large as 150 %, depending on the target. The mean energy of the sputtering events, $<E_s>$, shown in Fig. 3b, has a similar large dependence on $\theta$, and decreases with Z due to the larger stopping powers. The observed dependence of Y and $<E_s>$ on $\theta$ is in intuitive agreement with the analysis of the CDM flux anisotropy.

Fig. (4) further illustrates the effect. The differential rate of sputtering as a function of $E_s$ is shown for different values of $\theta$ in the case of a tellurium target sheet and a 100 GeV/$c^2$ Dirac neutrino. The figure shows the large daily fluctuation to be expected in this CDM signature, both in abundance and energy of the ejected atoms. Finally, Fig. (5) exemplifies the dependence of Y and $<E_s>$ on the mass of the WIMP, this time for a germanium target; information about the mass and coupling of the CDM particle would be gathered from a measurement of Y and $<E_s>$.

## 6. POSSIBLE METHODS OF DETECTION

In this section we propose two experimental techniques that could detect the daily modulations in Y and $<E_s>$, with the capability of yielding strong evidence for a CDM halo. Both are based on presently available technologies and hold the promise of near-perfect background rejection, necessary to discern the modulation.

The largest value of Y for Dirac neutrinos in the GeV - TeV mass range is ~ 10 sputtering events / $m^2$/ year, which is *prima facie* not too encouraging, especially since the expected neutralino signal starts two orders of magnitude below that value. However, the thickness of the target material in these searches does not need to exceed ~ 1000 Å, allowing the stacking of large surface areas in relatively small volumes. The general strategy is to alternate sheets of target material with layers of a detector sensitive to the energy of the sputtered atoms. The expected CDM signal in the detector material should be much smaller than in the target.

The first approach makes use of a time projection chamber (TPC) filled with planes of target material. This type of TPC consists of a gas chamber (at high or low pressure, depending on the application), containing a grid of anode and cathode wires close to its internal side walls. The grid produces an uniform electric field throughout the vessel, parallel to the target planes. Secondary electrons produced by ionization in the gas are drifted by the field to the grid, where charge collection takes place. Signals are electrostatically induced on a XY readout plane, and their time evolution provides information on the Z coordinate. The location of the event is thus specified. Typical spatial resolutions are of the order of 1 (mm)$^3$ or better. The energy of the event can be derived from integration of the charge collected over the drift time. In some applications, a magnetic field is applied across the chamber, forcing particles with different momenta and / or charge to trace trajectories of a different topology. Usual energy resolutions are ~ 10 % in either approach.

TPCs have been used extensively in a variety of searches [26] and most recently in double beta decay experiments [27,28,29], where their background rejection potential has been demonstrated. It is in some of these experiments that sheets of plastic (normally mylar) containing the source have been introduced in the active region of the chamber [28]. In one proposal [29] (currently under development) multiple sheets are present, with an active area of 2.5 m$^2$ in a sensitive volume of 60 x 60 x 120 cm$^3$. The samples must be thin in order to avoid deterioration of the energy resolution and the introduction of background. Preferably, the samples must be non-conducting, to avoid aberrations in the drift field. Special precautions should be taken if metallic samples must be used [29].

The simplified TPC that we propose measures energy from the charge collected (an applied magnetic field is not required), and contains a large density of sheets. Large spacing between sheets is only necessary when particle energies are derived from their trajectories in a magnetic field, which typically require many cm. The ranges of sputtered atoms in gas are much shorter. A modest volume of 1 m$^3$ can accommodate ~ 100 m$^2$ of target material without compromising the number of electronic channels necessary.

Double-beta decay TPCs normally "sandwich" the source material (in powder form) between mylar sheets a few microns thick. The energy loss of ßß electrons in the plastic is negligible. However, the range of even the heaviest atoms with energy < 100 keV is of less than 1 micron in a typical polymer. Single-side coating of the plastic with the selected target material is necessary instead for a CDM application. Shallow ion-implantation or any of the numerous industrial coating techniques [30] would be adequate for this purpose. The contribution to sputtering from the uncoated side is negligible even when the CDM "wind" favors it (i.e., when the coated side is at θ=180°). This is due to the low values of Y for the carbon, hydrogen and oxygen components of common polymers (Fig. 3a). We have examined this question in the case of mylar ($C_{10}H_8O_4$) coated with germanium; the contribution of mylar to total sputtering from the sheet varies between $10^{-1}$ ($\theta_{coated} = 180°$) and $10^{-3}$ ($\theta_{coated} = 0°$).

Even for thin plastic (~10 μm), recoiling nuclei going into the sheet can not reach the opposite side. This, together with the mentioned small contribution to Y from the plastic, guarantee that sputtered atoms are only ejected from the coated side, which should behave in accordance to the predictions of Figs. (3 a,b). The coated surface can be placed in the TPC facing in opposite directions from one sheet to the next. In this way, when some of the targets are at θ=0°, the rest are at θ=180°, increasing the uniqueness of the signature.

Not all the energy of the sputtered atoms is lost to ionization processes in the gas. The TPC is only sensitive to this fraction, $E_{ioniz}$. This results in "quenching", which must be accounted for when selecting the target material. We have analyzed the case of a helium TPC at 1 atm, where a modest electronic threshold of $E_{ioniz} > E_{thr} = 2$ keV has been imposed. For simplicity, no additives to the gas have been considered (20 % $CH_4$ is commonly used to increase the drift velocity). The electronic and nuclear losses in the gas for the different energetic ions are computed by TRIM. We calculate the number of events / $m^2$ / year with $E_{ioniz}$ above threshold for the specific case of a 100 GeV/$c^2$ Dirac neutrino (Fig. 6). The fraction of electronic energy loss is calculated independently for each atom sputtered in an event, and $E_{ioniz}$ is obtained from their sum. Heavy elements are disfavored under these experimental conditions, due to their lower $<E_s>$ (Fig. 3b) and higher nuclear stopping powers in the gas. Elements in the interval 30 < Z < 50 register a maximum of ~ 2 counts / $m^2$ / year above $E_{thr}$ (Fig. 6).

Relaxing the threshold to $E_{ioniz} > 1$ keV increases this figure to ~ 2.5 counts / m$^2$ / year for all elements with Z > 30.

The advantage of using a room-pressure TPC relies on the large difference in trajectory length for electrons and heavy ions of equivalent $E_{ioniz}$ (low-energy electrons lose most of their kinetic energy to ionization in the gas); Fig. 7 displays this difference in helium at 1 atm. Minimum ionizing radiation (gamma and x-rays, electrons, etc.) lose all their energy to secondary electrons. The CDM sputtering signature would consist of a single "spark" event at the position of one of the target sheets, singled-out in time and entirely contained within one of the spatial resolution 1 (mm)$^3$ "pixels" (lower right quadrant in Fig. 7). Electrons with an equivalent energy will "light up" more than one pixel (upper right quadrant), allowing near-perfect background rejection. The electron range has been calculated using the Bethe-Bloch expression [31] and is in agreement with Ref. [32]. We have estimated the electron straggling using the approximation to Moliere scattering in Ref. [33] and found it to be negligible above 3 keV. Only a fraction (~ 30 %) of the electrons with energies between 2 keV and 3 keV can straggle enough to lose all energy within the 1 (mm)$^3$ pixel.

Sputtering from neutron interactions in the plastic remains the single most important source of background in such a device. A conservative prediction, using the measured neutron flux at an underground installation [34] and allowing for a modest shielding, indicates that this process starts to compete with CDM sputtering only for WIMP couplings two to three orders of magnitude below that for Dirac neutrinos, depending on $m_\delta$. A TPC as described, with a target Z ~ 30, has a CDM sensitivity comparable to that of a typical ultra-low background germanium detector [19] after a data acquisition period of only ~ 30 m$^2$ - year. This sensitivity increases with the addition of new data.

In order to extensively probe the neutralino region, target surface areas of O($10^4$) m$^2$ are desirable. Certain types of solid state detectors may be suitable for this purpose. Metastable superconducting strips [35] stacked with target sheets could reach this limit. However, this type of detector is in an early stage of development and its response to radiation is currently under study. A more realistic possibility is based on solid state nuclear track detectors (SSNTDs) [36]. Heavily ionizing particles (α-particles, fission fragments, recoiling nuclei, etc.) produce tracks of solid-state damage (displaced lattice atoms, broken molecular bonds, etc.) in these materials. The damage remains latent for long periods of time in thermally stable samples. The tracks can later be revealed by etching with a chemical reagent. Essentially, etching occurs along a particle track faster than along the undamaged surface of the bulk material, due to the rapid dissolution of the damaged region. A surface etch-pit is thus formed and can be inspected by optical or electron microscopy, among other techniques. The main characteristic of interest in SSNTD materials is their insensitivity to minimum ionizing radiation, making them ideal for CDM searches [37].

The most sensitive of man-made SSNTDs is CR-39 plastic ($C_{12}O_7H_{18}$), whose response to low-velocity ions has been recently measured [38]. While the number of etchable tracks from CDM interactions in CR-39 itself is negligibly small, sputtered atoms from a target in direct contact with it may have an energy above the track registration threshold. Snowden-Ifft and Price [38] have measured the registration threshold and expressed it as a function of the stopping powers calculated by the TRIM code. We have used their prescription to obtain this threshold for perpendicularly incident ions of various Z (Fig. 8). A heavy atom (Z ~ 80) with as little as 2 keV of kinetic energy can leave an etchable track in CR-39. Also displayed in Fig. 8 is the number of etchable tracks induced by sputters from different targets in direct contact with the plastic, expressed in units of etch-pits / $m^2$/ year of exposure to the CDM flux (at a fixed θ). The calculation includes only those events above threshold. This density of pits is indeed an upper limit, since the registration threshold is higher for particles penetrating the CR-39 at a slant angle. However, the correction factor required is small; for Z ~ 80 the quasi-isotropy of the sputtered species decreases the expected density of etch-pits in Fig. (8) by only ~ 25 %.

In order to compensate for the lack of real-time event information in SSNTDs, active tracking of the direction of the CDM "wind" is required to observe the modulation; large surface areas of CR-39 and target film can be pressed together in rolls or stacks and placed on an X-Y table that keeps their orientation fixed with respect to **w**. Regions of the CR-39 maintained at different values of θ will register a different sputtering signal, as indicated in Fig. 8. The detector should be kept in vacuum during the exposure, to avoid energy loses in the few microns of air that might exist between the target and CR-39. The performance of CR-39 seems to be unaffected by vacuum conditions [38]. The primary experimental challenge is the automatization of the etching and later scanning of the plastic. Fortunately, extraordinary advances in this field have allowed the fast processing of $4 \times 10^3$ m$^2$ of CR-39 in a search for magnetic monopoles [39]. Fully automated scanning and identification of even very shallow pits that merge asymptotically with the plastic surface has been accomplished recently [40]. The physical characteristics of the etch-pit convey information about the nature and energy of the causing particle, and are measured with a precision that surpasses that of a skilled human observer. State of the art automated systems can discriminate true tracks from nontrack surface defects (whose density is very small in CR-39) even when their ratio is as low as 1 / 1000 [40].

## 7. CONCLUSIONS

Any interesting direct-detection experiment must aim at the observation of an unambiguous signature of the CDM signal. The advantages of a large modulation of diurnal periodicity combined with an efficient background rejection, are obvious. We have presented one such signature, WIMP-induced surface sputtering, and calculated its daily variation in a variety of materials, with emphasis on two actual experimental designs. These seem to be immediately feasible, but not exclusive. The present calculations are intended to be applied to the design of experiments that exploit this promising technique.

**ACKNOWLEDGMENTS**

We are indebted to J. F. Ziegler at IBM-Research for making the TRIM code available to us. This work was supported in part by the National Science Foundation under Grant No. PHY-9210924.

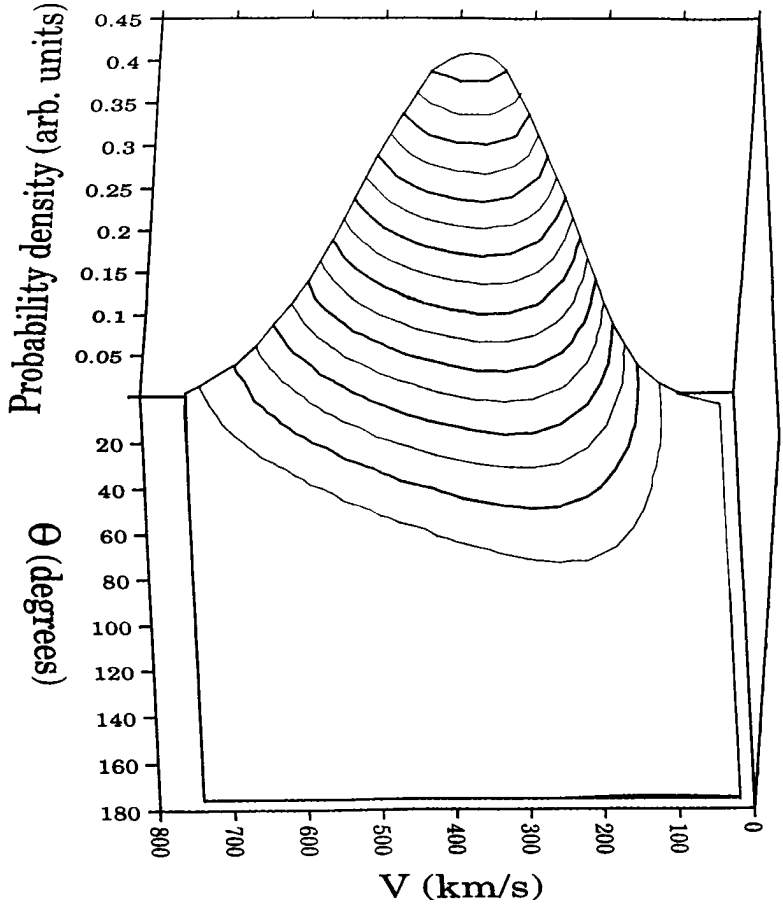

Fig. 1

Probability distribution of CDM velocity vectors in the Earth's reference frame. The polar angle θ is measured from the azymuthal symmetry axis defined by $\mathbf{w} = -\mathbf{v}_{Earth}$, the negative of the velocity of the Earth through the galactic halo. The relevant speeds assumed here are $v_{Earth} = 260\,km/s$, $v_{dis} = 270\,km/s$, $v_{esc} = 550\,km/s$ (see text).

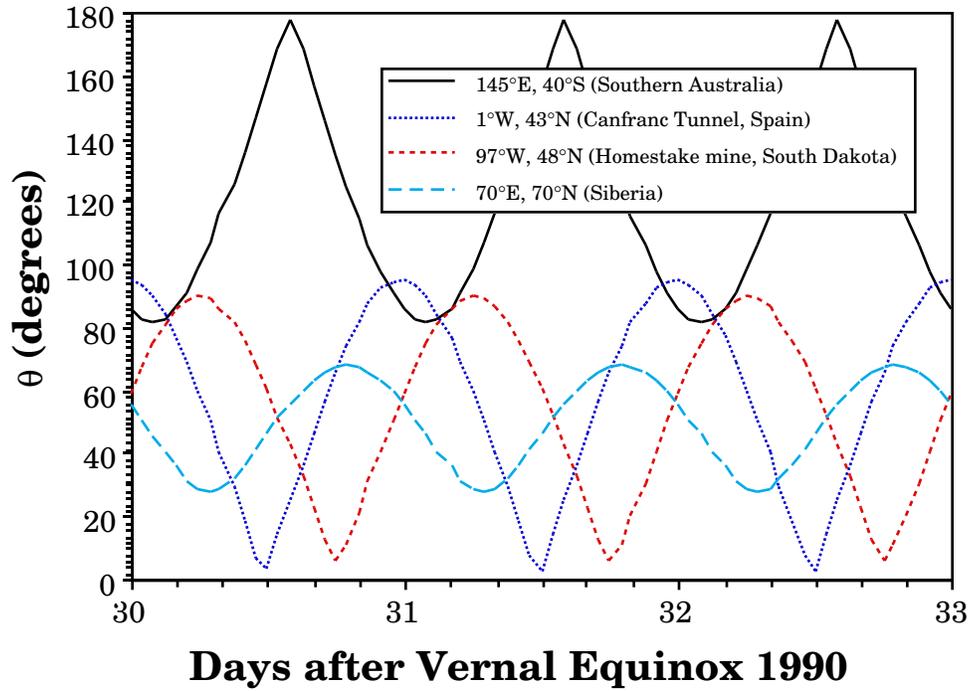

Fig. 2

Daily changes in θ for the vector perpendicular to a sheet of target material. The vector is oriented in the negative vertical direction (towards the center of the Earth). The strong dependence on the geographical location of the detector is evident.

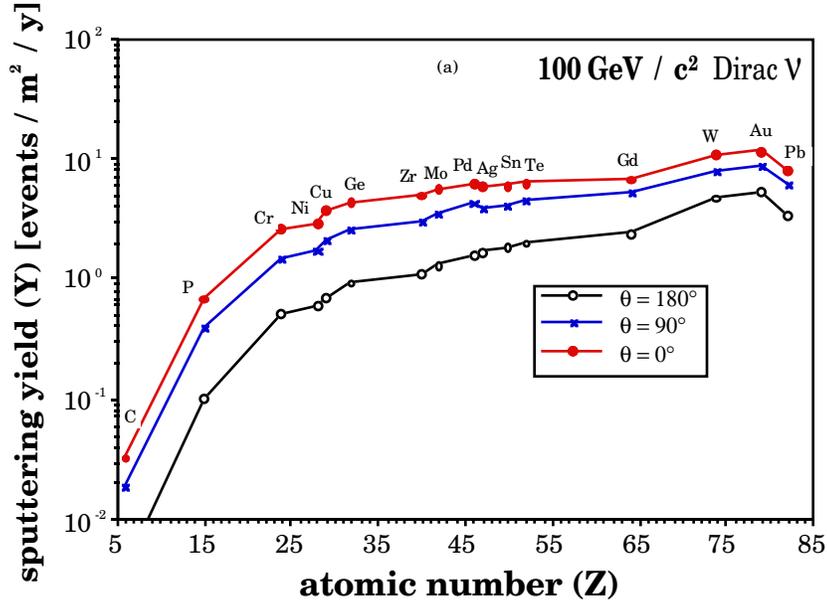
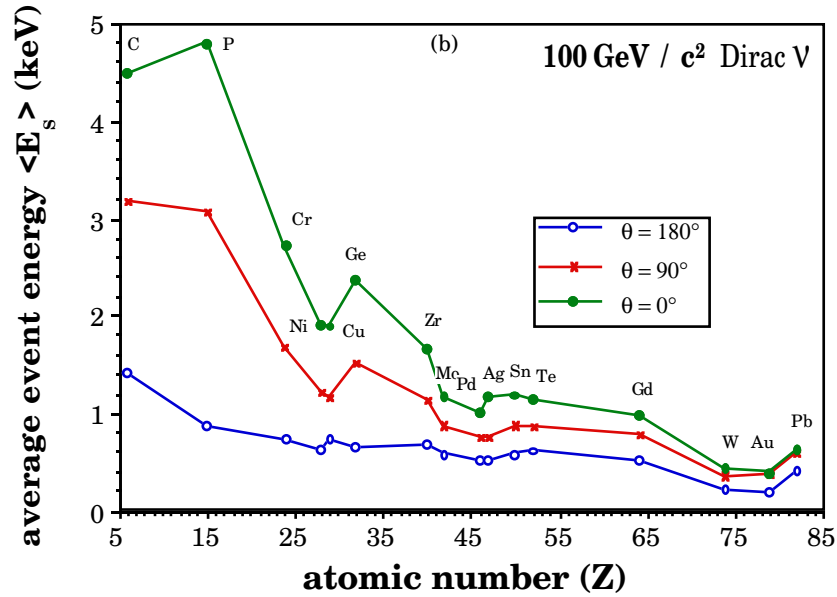

Figs. 3a,b

Sputtering yield in different elements from elastic scattering of a reference CDM particle ( $100$ GeV/$c^2$ Dirac neutrino). The local halo density used is $\rho_{halo} = 0.4$ GeV/$c^2$/cm$^3$. The angle $\theta$ defines the orientation of the target material with respect to the galactic "wind". The rotation of the Earth changes $\theta$ for an otherwise stationary target, creating a strong diurnal variation in the sputtering signal. Fig. 3b displays the dependence of the mean energy of the sputtered species on atomic number and target orientation.

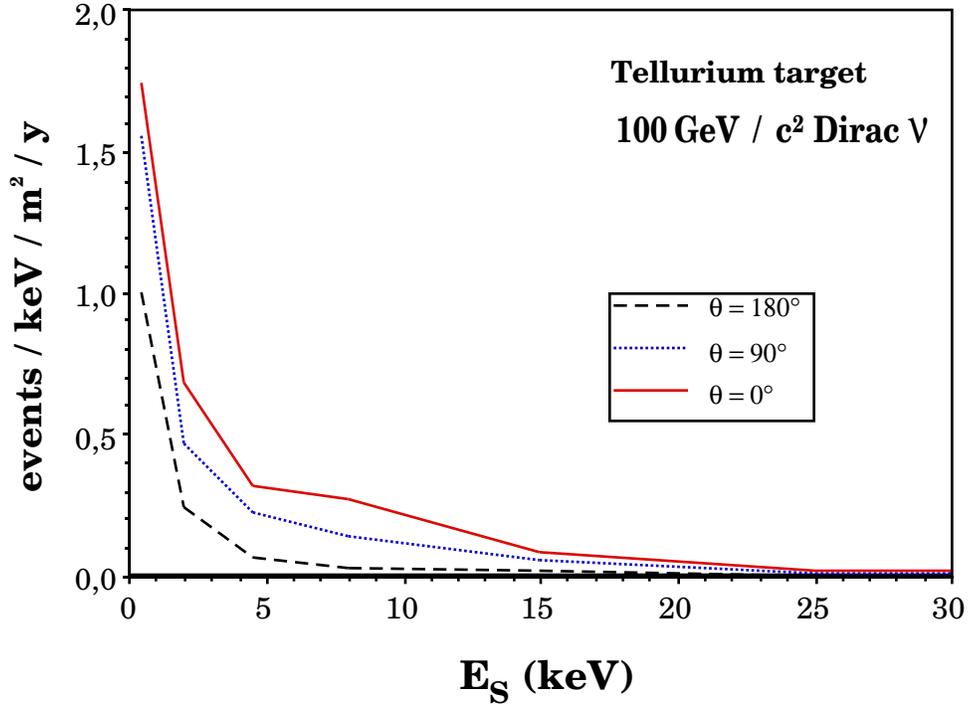

Fig. 4

Differential rates of sputtering for a tellurium target and a $100~\text{GeV}/c^2$ Dirac neutrino galactic halo. This particle is rejected by underground germanium experiments as the main constituent of the halo, and is used here only to illustrate the magnitude of the modulation in the sputtering signal. For the still allowed sub-$Z^0$ couplings, these sputtering rates scale as $(G_f/G_w)^2$ (Eq. 6).

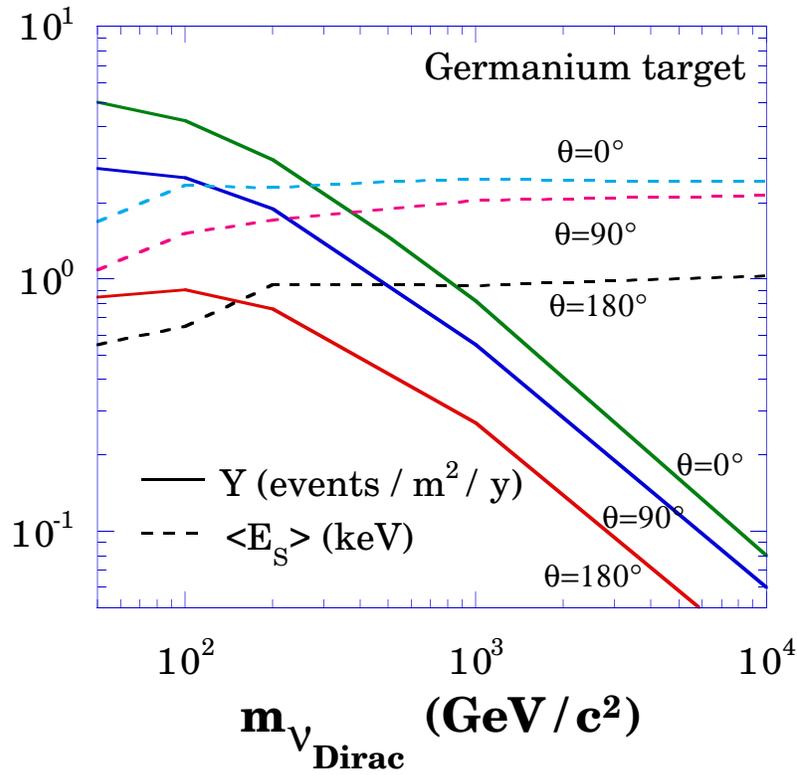

Fig. 5

Sputtering yield and mean energy of the ejected atoms from a germanium target, as a function of the mass of the Dirac neutrino assumed to constitute the galactic halo. These values can be scaled to cover other CDM particles, including some neutralino candidates (see text).

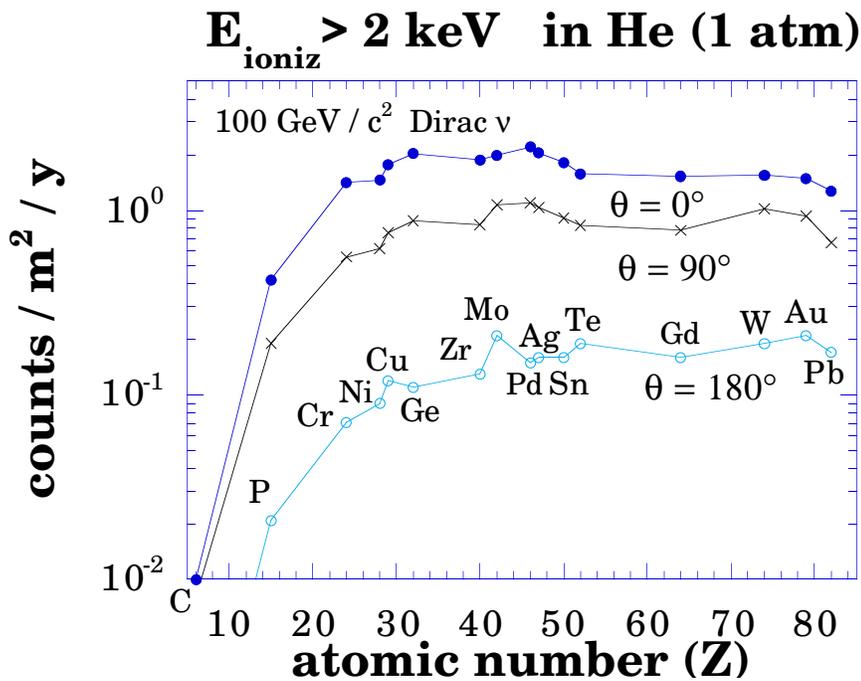

Fig. 6

Predicted modulation in the rate of sputtering events for the time projection chamber described in the text (helium at 1 atm), above an electronic threshold of $E_{ioniz} > 2$ keV. Target elements in the region $30 < Z < 50$ are favored in this kind of search for CDM particles of mass $\sim 100$ GeV/$c^2$.

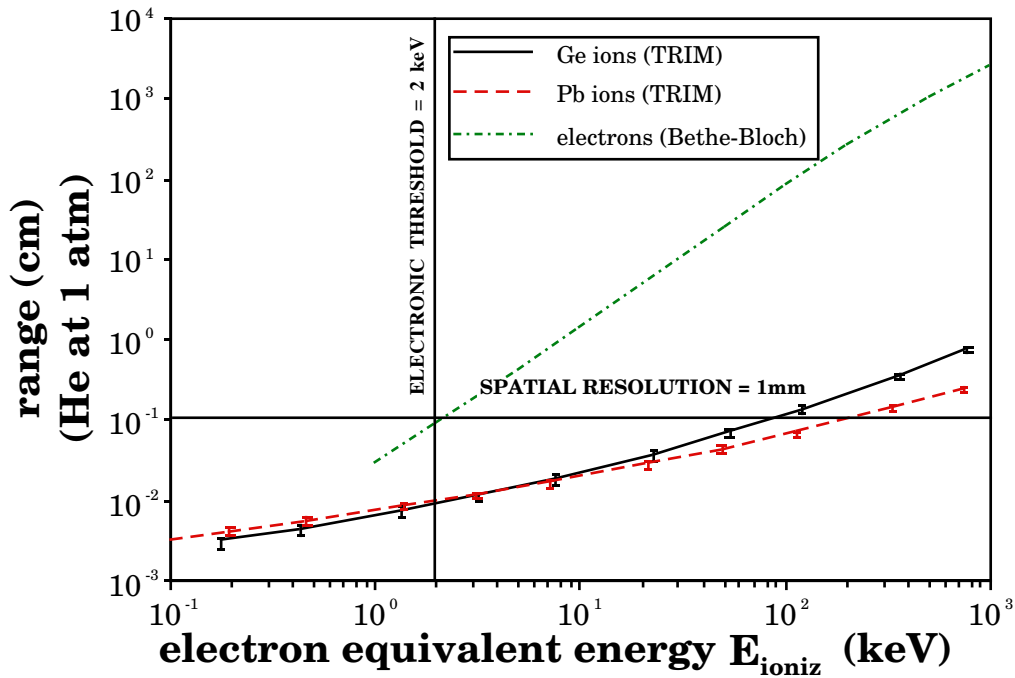

Fig. 7

Projected range for electrons and heavy ions in a helium TPC (1 atm). The error bars represent the longitudinal straggling of the ions, obtained from TRIM. Electrons with $E_{ioniz}$ above 2 keV can be rejected with a typical spatial resolution of 1 mm$^3$, with the consequent reduction in the background associated to minimum ionizing radiation.

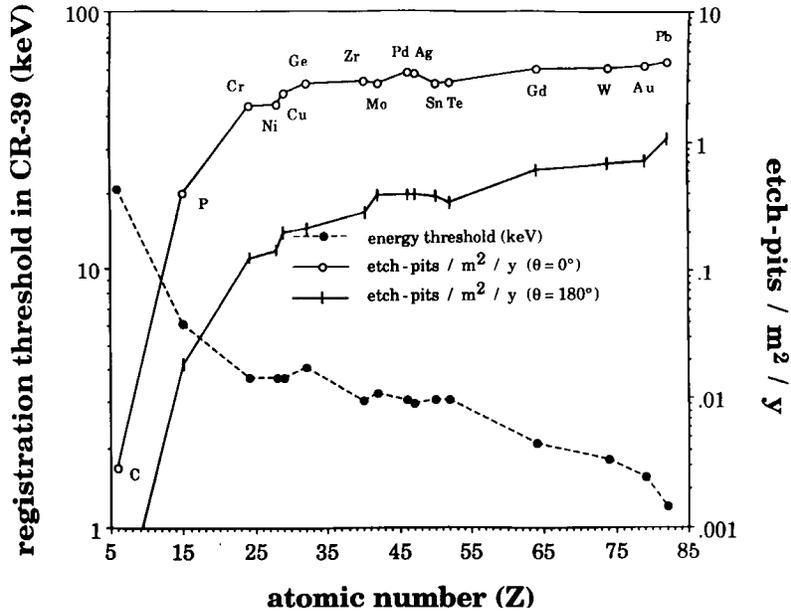

Fig. 8

Minimum kinetic energy necessary for the formation of an etchable track in CR-39 by an ion perpendicularly penetrating the plastic. The calibration is taken from the measurements of Ref. [38]. Using this registration threshold, we obtain the approximate number of etch-pits expected in a sheet of CR-39 in direct contact with a target surface maintained at $\theta=0°$ and $\theta=180°$ (see text). The projectile CDM particle considered is a 100 GeV/$c^2$ Dirac neutrino.